\documentclass[aps,pre]{revtex4}
\usepackage{graphicx}
\usepackage{dcolumn}% Align table columns on decimal point
\usepackage{bm}% bold math

\begin{document}
\title{
Solving the Maxwell equations by the Chebyshev method:\\
A one-step finite-difference time-domain algorithm}
\author{%
H. De Raedt\footnote{E-mail: deraedt@phys.rug.nl},
K. Michielsen\footnote{E-mail: kristel@phys.rug.nl},
J.S. Kole\footnote{E-mail: j.s.kole@phys.rug.nl},
M.T. Figge\footnote{E-mail: m.t.figge@phys.rug.nl\\
http://www.compphys.rug.nl/}
}
\affiliation{%
Centre for Theoretical Physics and Materials Science Centre\\
University of Groningen, Nijenborgh 4\\
NL-9747 AG Groningen, The Netherlands
}
\date{\today}
\begin{abstract}%
We present a one-step algorithm that solves the Maxwell equations for systems
with spatially varying permittivity and permeability by the Chebyshev method.
We demonstrate that this algorithm may be orders of magnitude more efficient than
current finite-difference time-domain algorithms.
%We consider a one-dimensional system with a current source and compute the
%frequency spectrum of a three-dimensional photonic crystal with a woodpile
%structure to demonstrate the virtues of the one-step algorithm.
\end{abstract}


\maketitle

%%%%%%%%%%%%%%%%%%%%%%%%%%%%%%%%%%%%%%%%%%%%%%%%%%%%%%%%%%%%%%%%%%%%%%%%%%%%
%
\newcommand{\threevec}[3]{\left(\begin{array}{c}%
 #1 \\ #2 \\ #3 \end{array}\right)}
\newcommand{\ve}{\varepsilon}
\newcommand{\sve}{\sqrt{\varepsilon}}
\newcommand{\smu}{\sqrt{\mu}}
\def\bE{{\mathbf{E}}}
\def\bH{{\mathbf{H}}}
\def\bD{{\mathbf{D}}}
\def\bB{{\mathbf{B}}}
\def\bX{{\mathbf{X}}}
\def\bY{{\mathbf{Y}}}
\def\bZ{{\mathbf{Z}}}
\def\bJ{{\mathbf{J}}}
\def\bj{{\mathbf{s}}}
\def\bk{{\mathbf{k}}}
\def\be{{\mathbf{e}}}
\def\br{{\mathbf{r}}}
\def\bPsi{{\mathbf{\Psi}}}
\def\bPhi{{\mathbf{\Phi}}}
\def\bsigma{{\mathbf{\sigma}}}
\def\bXi{{\mathbf{\Xi}}}
\def\bzero{{\mathbf{0}}}
\newcommand{\php}{\phantom{+}}
\newcommand{\pht}{\phantom{T}}
\newcommand{\dd}[1]{\frac{\partial}{\partial #1}}
\section{Introduction}\label{sec1}

Most finite-difference time-domain (FDTD) calculations
solve the time-dependent Maxwell equations using algorithms
based on a proposal by Yee~\cite{Taflove,Kunz,Yee66}.
The Yee algorithm is flexible, fast and easy to implement.
A limitation of Yee-based FDTD techniques is that their stability is
conditional, meaning that their numerical stability
depends on the mesh size used for the spatial discretization
and on the time step of the time integration~\cite{Taflove,Kunz}.
%
%Recently a family of unconditionally stable algorithms has been introduced~\cite{Kole01,Kole02}.
%The strategy behind this approach is
%to approximate the formal solution of the time evolution of the electromagnetic (EM) fields,
%which can be written as the exponential of a skew-symmetric matrix~\cite{Kole01,Kole02},
%in terms of orthogonal transformations. The resulting numerical algorithms are
%unconditionally stable by construction~\cite{DeRaedt87}.
In practice, the amount of computational work required to solve the time-dependent Maxwell equations
by present FDTD techniques~\cite{Taflove,Kunz,Website,Zheng1,Namiki,Zheng2,Harsh00,Kole01,Kole02}
prohibits applications to a class of important fields such as
bioelectromagnetics and VLSI design~\cite{Taflove,Gandi,Houshmand}.
The basic reason for this is that the time step in the FDTD calculation has to be relatively small in
order to maintain a reasonable degree of accuracy in the time integration.

In this paper we describe a one-step algorithm,
based on Chebyshev polynomial expansions~\cite{TAL-EZER0,TAL-EZER,LEFOR,Iitaka01,SILVER,LOH},
to solve the time-dependent Maxwell equations for arbitrarily long times.
We demonstrate that the computational efficiency of this one-step algorithm can be orders of magnitude
larger than of other FDTD techniques.

\section{Algorithm}\label{sec2}

We consider EM fields in linear, isotropic, nondispersive and lossless materials.
The time evolution of EM fields in these systems is governed by the time-dependent
Maxwell equations~\cite{BornWolf}.
Some important physical symmetries of the Maxwell equations can be made explicit by introducing
the fields
\begin{equation}
\bX(t)\equiv\sqrt{\mu}\,\bH(t)\quad {\rm and} \quad
\bY(t)\equiv\sqrt{\varepsilon}\,\bE(t)\,. \label{EY}
\end{equation}
Here, $\bH(t)=(H_x(\br,t),H_y(\br,t),H_z(\br,t))^T$ denotes the magnetic
and $\bE(t)=(E_x(\br,t),E_y(\br,t),E_z(\br,t))^T$
the electric field vector, while $\mu=\mu(\br)$ and $\ve=\ve(\br)$ denote,
respectively, the permeability and the permittivity.
In the absence of electric charges, Maxwell's curl equations~\cite{Taflove}
read
\begin{equation}
\dd{t}
\left(
\begin{array}{c}
\bX(t) \\
\bY(t)
\end{array}
\right)
={\mathcal H}
\left(\begin{array}{c} \bX(t) \\\bY(t) \end{array} \right)
- \frac{1}{\sqrt{\ve}} \left(\begin{array}{c} 0 \\ \bJ(t)   \end{array} \right),
\label{TDMEJ}
\end{equation}
where $\bJ=(J_x(\br,t),J_y(\br,t),J_z(\br,t))^T$ represents the source of
the electric field and $\mathcal H$ denotes the operator
\begin{equation}
{\mathcal H}  \equiv
\left( \begin{array}{cc} 0 &
  -\frac{1}{\sqrt{\mu}}\mathbf{\nabla}\times\frac{1}
  {\sqrt{\varepsilon}} \\
   \phantom{-}\frac{1}{\sqrt{\varepsilon}}\mathbf{\nabla}
   \times\frac{1}{\sqrt{\mu}} & 0
 \end{array}\right).
\label{Hconti}
\end{equation}
Writing $\bZ(t)=(\bX(t),\bY(t))^T$ it is easy to show that ${\mathcal H}$
is skew symmetric, i.e. ${\mathcal H}^T=-{\mathcal H}$, with respect to the
inner product $\langle\bZ|\bZ^\prime\rangle
\equiv\int_V\bZ^T\cdot\bZ^\prime\, d\br$, where $V$ denotes the system's
volume.
In addition to Eq.(\ref{TDMEJ}), the EM fields also satisfy
$\nabla\cdot(\sqrt{\mu}\bX(t)) = 0$ and
$\nabla\cdot(\sqrt{\ve}\bY(t)) = 0$.

A numerical algorithm that solves the time-dependent Maxwell equations
necessarily involves some discretization procedure of the spatial derivatives
in Eq.~(\ref{TDMEJ}).
Ideally, this procedure should not change the basic symmetries of the
Maxwell equations.
We will not discuss the (important) technicalities of the
spatial discretization (we refer the reader to Refs.~\cite{Kole01,Kole02})
as this is not essential to the construction of the one-step algorithm.

On a spatial grid Maxwell's curl equations (\ref{TDMEJ})
can be written in the compact form~\cite{Kole01,Kole02}
\begin{equation}
\dd{t}\bPsi(t)=H\bPsi(t)-\bPhi(t)\,.
\label{PsiJ}
\end{equation}
The vector $\bPsi(t)$ is a representation of $\bZ(t)$
on the grid. The matrix $H$ is the discrete analogue of the
operator (\ref{Hconti}), and the vector $\bPhi(t)$ contains all the
information on the current source $\bJ$.
The formal solution of Eq.~(\ref{PsiJ}) is given by
\begin{equation}
\bPsi(t)=U(t)\bPsi(0)-\int_0^{t}U(t-u)\bPhi(u)du\,,
\label{FormalJ}
\end{equation}
where $U(t)=e^{tH}$ denotes the time-evolution matrix.
The underlying physical symmetries of the time-dependent
Maxwell equations are reflected by the fact that the matrix $H$ is real and skew
symmetric~\cite{Kole01}, implying that $U(t)$ is orthogonal~\cite{WILKINSON}.

\begin{figure}[t]
\begin{center}
\includegraphics[width=16.cm]{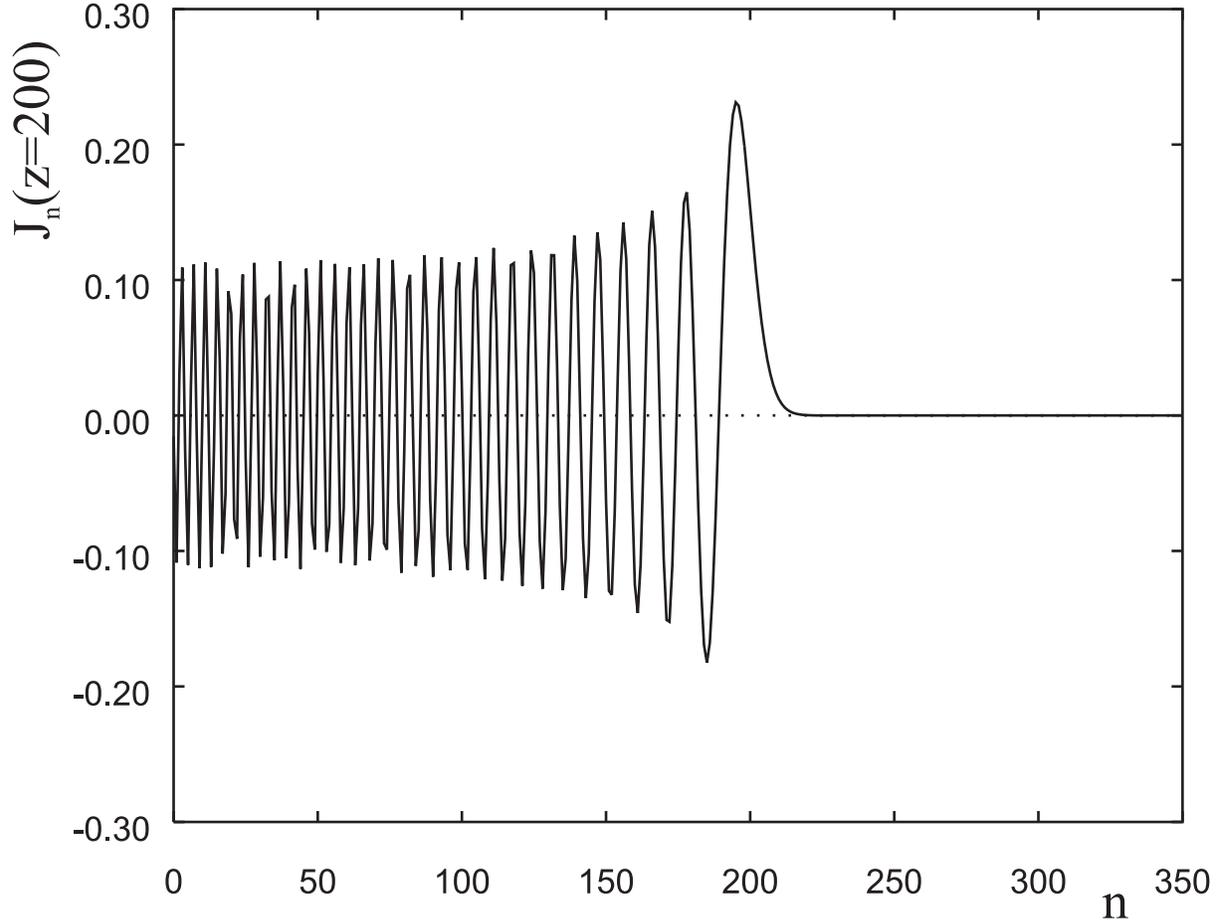}
\caption{Dependence of the Bessel function $J_n(z=200)$ on the order $n$.}
\label{fig:Jn}
\end{center}
\end{figure}

Numerically, the time integration is carried out by using a time-evolution operator $\overline{U}(t)$
that is an approximation to $U(t)=e^{tH}$.
We denote the approximate solution by $\overline{\bPsi}(t)$.
First we use the Chebyshev polynomial expansion
to approximate $U(t)$ and then show how to treat the source term
in Eq.~(\ref{FormalJ}).
We begin by ``normalizing'' the matrix $H$.
The eigenvalues of the skew-symmetric matrix $H$ are pure imaginary numbers.
In practice $H$ is sparse so it is easy to compute $\Vert H\Vert_1\equiv\max_j \sum_i |H_{i,j}|$.
Then, by construction, the eigenvalues of $B\equiv -iH/\Vert H\Vert_1$
all lie in the interval $[-1,1]$~\cite{WILKINSON}.
Expanding the initial value $\bPsi(0)$ in the (unknown) eigenvectors ${\bf b}_j$ of
$B$, we find from Eq.~(\ref{FormalJ}) with $\bPhi(t)\equiv 0$:
\begin{equation}
\bPsi(t)=
e^{izB}\bPsi(0) =\sum_j e^{izb_j} {\bf b}_j
\langle{\bf b}_j|\bPsi(0)\rangle,
\label{expz}
\end{equation}
where the $b_j$ denote the (unknown) eigenvalues of $B$.
Although there is no need to know the eigenvalues and eigenvectors of $B$ explicitly,
the current mathematical justification of the Chebyshev approach requires that
$B$ is diagonalizable and that its eigenvalues are real. The effect of relaxing these conditions
on the applicability of the Chebyshev approach is left for future research.
We find the Chebyshev polynomial expansion of $U(t)$ by computing the
expansion coefficients of each of the functions $e^{izb_j}$ that appear
in Eq.~(\ref{expz}).
In particular, as $-1\le b_j \le 1$, we can use the expansion~\cite{ABRAMOWITZ}
$
e^{izb_j}=J_0(z) + 2\sum_{k=1}^{\infty} i^{k}J_{k}(z)T_{k}(b_j)\,,
$
where $J_k(z)$ is the Bessel function of integer order $k$
to write Eq.~(\ref{expz}) as
\begin{equation}
\bPsi(t)
=\left[J_0(z)I + 2\sum_{k=1}^{\infty} J_{k}(z)\widetilde T_{k}(B)\right] \bPsi(0)\,.
\label{SUM0}
\end{equation}
Here, $I$ is the identity matrix and $\widetilde T_{k}(B)=i^k T_k(B)$ is a
matrix-valued modified Chebyshev polynomial that is defined by the recursion
relations
\begin{equation}
\widetilde T_{0}(B)\bPsi(0)=\bPsi(0)\label{CHEB1}\,,\quad
\widetilde T_{1}(B)\bPsi(0)=iB\bPsi(0)\,,
\label{CHEB44}
\end{equation}
and
\begin{equation}
\widetilde T_{k+1}(B)\bPsi(0)=
2iB\widetilde T_{k}(B)\bPsi(0)+\widetilde T_{k-1}(B)\bPsi(0)\,,
\label{CHEB4}
\end{equation}
for $k\ge1$.
In practice we truncate the sum in Eq.~(\ref{SUM0}), i.e.
to obtain the approximation $\overline{\bPsi}(t)$ we will sum only
the contributions with $k\leq K$.
As $\Vert \widetilde T_{k}(B) \Vert_1\le1$ by construction and
$|J_k(z)|\le |z|^k/2^k k!$ for $z$ real~\cite{ABRAMOWITZ},
the resulting error vanishes exponentially fast for sufficiently large $K$.
In Fig.\ref{fig:Jn} we show a plot of $J_n(z=200)$ as a function of $n$
to illustrate this point.
From Fig.\ref{fig:Jn} it is clear that the Chebyshev polynomial expansion will only be useful if
$K$ lies to the right of the right-most extremum of $J_n(z=200)$.
From numerical analysis it is known that for fixed $K$, the Chebyshev polynomial is very nearly
the same polynomial as the minimax polynomial \cite{NumercalRecipes}, i.e.
the polynomial of degree $K$ that has the smallest maximum deviation from the true function,
and is much more accurate than for instance a Taylor expansion of the same degree $K$.
The coefficients $J_{k}(z)$ should be calculated to high precision
and the number $K$ is fixed by requiring that $|J_{k}(z)|<\kappa$ for all $k>K$.
Here, $\kappa$ is a control parameter that determines the accuracy of the approximation.
For fixed $\kappa$, $K$ increases linearly with $z=t\Vert H\Vert_1$ (there is
no requirement on $t$ being small),
a result that is essential for the efficiency of the algorithm.
Using the recursion relation of the Bessel functions, all $K$ coefficients
can be obtained with $\cal{O}(K)$ arithmetic operations \cite{NumercalRecipes}.
Clearly this is a neglible contribution to the total computational cost
for solving the Maxwell equations.

Performing one time step amounts to repeatedly using recursion (\ref{CHEB4}) to
obtain $\widetilde T_{k}(B)\bPsi(0)$ for $k=2,\ldots,K$, multiply the elements
of this vector by $J_{k}(z)$ and add all contributions.
This procedure requires storage for two vectors of the same length as $\bPsi(0)$
and some code to multiply such a vector by the sparse matrix $H$.
The result of performing one time step yields the solution at time $t$, hence
the name one-step algorithm.
In contrast to what Eqs.~(\ref{CHEB44}) and ~(\ref{CHEB4}) might suggest,
the algorithm does not require the use of complex arithmetic.

We now turn to the treatment of the current source $\bJ(t)$.
The contribution of the source term to the EM field at time $t$ is
given by the last term in Eq.~(\ref{FormalJ}).
One approach might be to use the Chebyshev expansion (\ref{SUM0}) for
$U(t-u)=e^{(t-u)H}$ and to perform the integral in Eq.~(\ref{FormalJ}) numerically.
However that is not efficient as for each value of $t-u$ we
would have to perform a recursion of the kind Eq.~(\ref{CHEB4}).
Thus, it is better to adopt another strategy.
For simplicity we only consider the case of a sinusoidal source
\begin{equation}
\bJ(\br,t)=\Theta(T-t)\bj(\br)\sin(\Omega t),
\label{SOURCE}
\end{equation}
where $\bj(\br)$ specifies the spatial distribution and $\Omega$ the angular
frequency of the source. The step function $\Theta(T-t)$ indicates that
the source is turned on at $t=0$ and is switched off at $t=T$.
%Artifacts that result from the discontinuity at $t=T$ can be miminized by
%chosing $T$ such that $T\Omega/2\pi$ is an integer number.
Note that Eq.~(\ref{SOURCE}) may be used to compose sources with a more
complicated time dependence by a Fourier sine transformation.

The formal solution for the contribution of the sinusoidal source (\ref{SOURCE}) reads
\begin{eqnarray}
\int_0^t e^{(t-u){H}}\bPhi(u)\,du&=&
(\Omega^2+H^2)^{-1} e^{(t-T^\prime)H}
\times(\Omega e^{T^\prime H}-\Omega\cos\Omega T^\prime -H\sin\Omega T^\prime)\bXi \nonumber \\
&\equiv& f(H,t,T^\prime,\Omega)\bXi\,,
\label{FORMALC}
\end{eqnarray}
where $T^\prime=\min(t,T)$ and $\bPhi(u)\equiv\Theta(T-t)\sin(\Omega t)\bXi$
with $\bXi$ a vector of the same length as $\bPsi(0)$ that represents the
time-independent, spatial distribution $\bj(\br)$.
The coefficients of the Chebyshev polynomial expansion of
the formal solution (\ref{FORMALC}) are calculated as follows.
First we repeat the scaling procedure described above and substitute in
Eq.~(\ref{FORMALC}) $H=ix\Vert H\Vert_1$, $t=z/\Vert H\Vert_1$,
$T^\prime=Z^\prime/\Vert H\Vert_1$, and $\Omega=\omega\Vert H\Vert_1$.
Then, we compute the (Fast) Fourier Transform with respect to $x$ of the function
$f(x,z,Z^\prime,\omega)$ (which is non-singular on the interval $-1\le x \le 1$).
By construction, the Fourier coefficients $S_k(t{\Vert H\Vert_1})$ are
the coefficients of the Chebyshev polynomial expansion~\cite{ABRAMOWITZ}.

Taking into account all contributions of the source term with $k$ smaller than
$K^\prime$ (determined by a procedure similar to the one for $K$),
the one-step algorithm to compute the EM fields at time $t$ reads
\begin{eqnarray}
\overline{\bPsi}(t)&=&
\left[J_0(t{\Vert H\Vert_1})I +
2\sum_{k=1}^{K} J_{k}(t{\Vert H\Vert_1})\widetilde T_{k}(B)\right] \bPsi(0)\nonumber\\
&+&\left[S_0(t{\Vert H\Vert_1})I + 2\sum_{k=1}^{K^\prime} S_{k}(t{\Vert H\Vert_1})
\widetilde T_{k}(B)\right] \bXi\,.
\label{APPROXJ}
\end{eqnarray}
We emphasize that in our one-step approach the time dependence of
the source is taken into account exactly, without actually sampling it.

\section{Results}\label{sec3}
The following two examples illustrate the efficiency of the one-step algorithm.
First we consider a system in vacuum ($\ve=\ve_0$ and $\mu=\mu_0$) which is
infinitely large in the $y$- and $z$-direction, hence effectively one dimensional.
The current source (\ref{SOURCE}) is placed at the center of a system of length 250.1 and
oscillates with angular frequency $\Omega=2\pi$ during the time interval $0\le t\le T=4$~\cite{units}.
In Table~\ref{tab1} we present results of numerical experiments with two
different time-integration algorithms.
%We compare the solutions of the Maxwell equations after simulation time $t=100$.
In general, the error of a solution $\widetilde\bPsi(t)$ as obtained by the
FDTD algorithm of Yee~\cite{Taflove,Yee66} or the unconditionally stable FDTD
algorithm $T4S2$~\cite{Kole01,Kole02} is defined by
$\Delta(t)\equiv
\Vert\widetilde\bPsi(t) - \overline{\bPsi}(t) \Vert/\Vert\overline{\bPsi}(t) \Vert$,
where $\overline{\bPsi}(t)$ denotes the vector of the EM fields as obtained by the one-step algorithm.
The error on the Yee-algorithm result
vanishes as $\tau^2$ for sufficiently small $\tau$~\cite{Taflove,Yee66}.
However, as Table I shows, unless $\tau$ is made sufficiently small ($\tau\le0.0125$ in this example),
the presence of the source term changes the quadratic behavior to almost linear.
The rigorous bound on the error between the exact and $T4S2$ results
tells us that this error should vanish as $\tau^4$~\cite{Kole01,DeRaedt87}.
This knowledge can be exploited to test if the one-step algorithm yields the
exact numerical answer.
Using the triangle inequality we can write
\begin{eqnarray}
\Vert\bPsi(t) - \overline{\bPsi}(t) \Vert
\;\le\;
\Vert\bPsi(t) - \widetilde\bPsi(t) \Vert &+&\Vert\widetilde\bPsi(t) - \overline{\bPsi}(t) \Vert
\nonumber \\
\le\;
\tau^4 t C \left(1+\int_0^t\Vert \bJ(u)\Vert du\right)
 &+& \Delta(t)  \Vert\overline{\bPsi}(t) \Vert
%\Vert\widetilde\bPsi(t) - \overline{\bPsi}(t) \Vert,
\label{ineq}
\end{eqnarray}
where $C$ is a positive constant~\cite{DeRaedt87}.
The numerical data in Table~\ref{tab1} (third column) show that
$\Delta(t)\rightarrow 0$
as $\tau^4$ and, therefore, we can be confident that the one-step algorithm
yields the correct answer within rounding errors.
Furthermore, since the results of the one-step algorithm are exact within almost machine precision,
in general the solution also satisfies $\nabla\cdot(\sqrt{\mu}\bX(t)) = 0$ and
$\nabla\cdot(\sqrt{\ve}\bY(t)) = 0$ within the same precision.
This high precision also allows us to use the one-step algorithm for genuine time stepping
with arbitrarily large time steps, this in spite of the fact that strictly speaking,
the one-step algorithm is not unconditionally stable.

From Table~\ref{tab1} it follows that if one finds
an error of more than 2.5\% acceptable, one could use the Yee algorithm, though we recommend to
use the one-step algorithm because then the time-integration error is neglegible.
The Yee algorithm is no competition for the $T4S2$ algorithm if one requires an error of
less than 1\%, but the $T4S2$ algorithm is not nearly as efficient as the one-step
algorithm with respect to the number of required matrix-vector operations.

\begin{figure}[t]
\begin{center}
\includegraphics[width=16.cm]{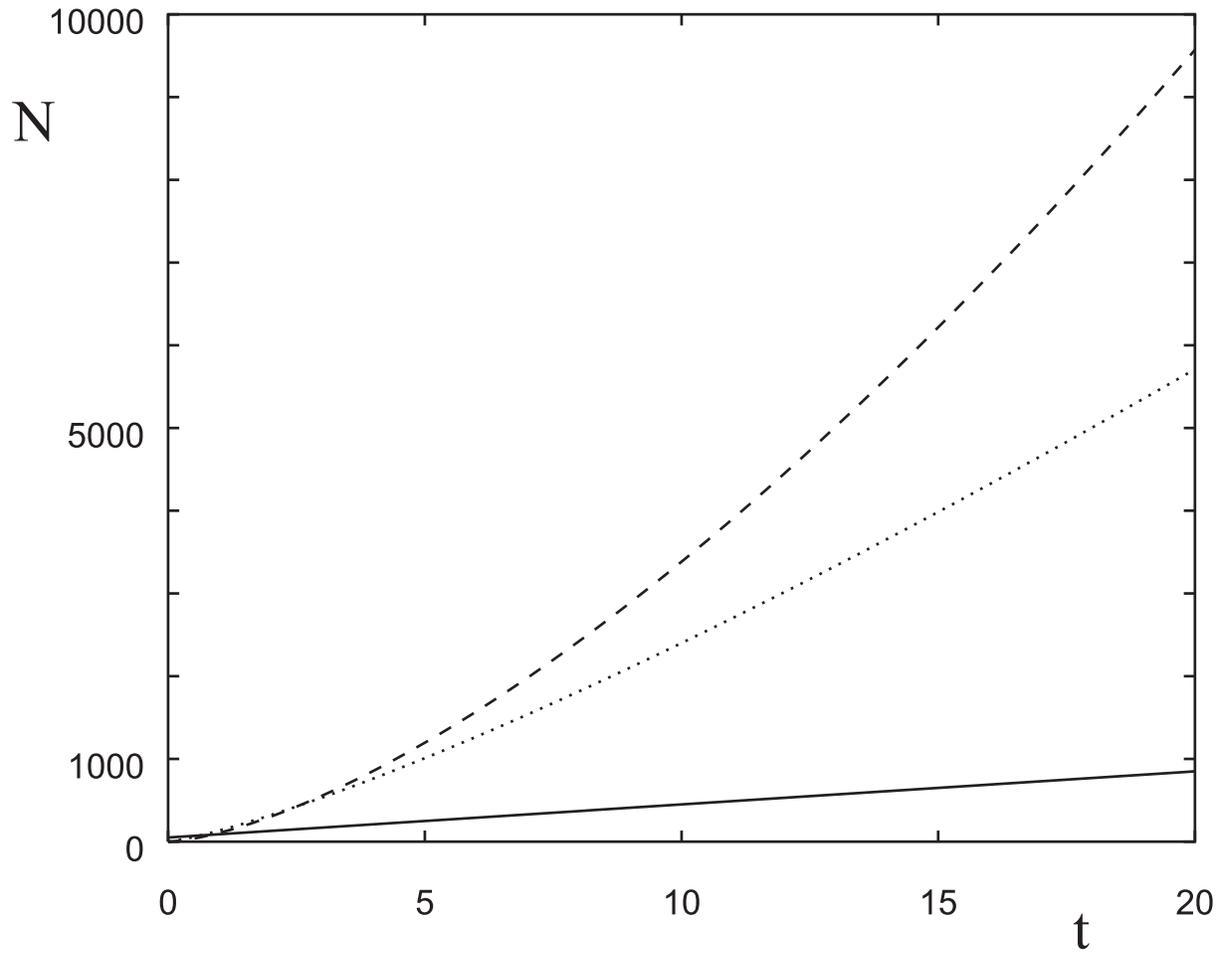}
\caption{The number of $\bPsi^\prime\leftarrow M\bPsi$ operations $N$ needed to compute the
solution of the 3D Maxwell equation at time $t$ for systems like those shown in Fig.\ref{woodpile}.
Solid line: One-step algorithm;
dashed line: Yee algorithm~\cite{Yee66,Taflove,Kunz} yielding a solution within 0.1\% error;
dotted line: T4S2 algorithm~\cite{Kole01,Kole02} yielding a solution within 0.1\% error.
}
\label{scalingplot}
\end{center}
\end{figure}

A more general quantitative analysis of the efficiency can be made using the fact
that for an $n$th-order algorithm ($n=2$ for the Yee algorithm and $n=4$ for the $T4S2$ algorithm),
the error $\Delta(t)$ vanishes no faster with $\tau$ than $\tau^n t$.
Each time step takes a number $W(n)$ of matrix-vector operations (of the type
$\bPsi^\prime\leftarrow M\bPsi$), e.g. for a
three-dimensional system we have $W(2)=1$ and $W(4)=10$ for the Yee algorithm and
the $T4S2$ algorithm, respectively.
In practice the actual number of floating point operations carried out by our
algorithms agrees with these estimates.
The total number of matrix-vector operations it takes to obtain the solution
at a reference time $t_r$ with error $\Delta_r(t_r)$ is then given by
$N_r=W(n) t_r/ \tau_r$ and thus $\Delta_r(t_r)\propto W(n)^n t_r^{n+1}/ N_r^n$.
The number of operations $N$ that it will take to compute the EM fields at time $t$
with accuracy $\Delta(t)$ is then calculated from
\begin{equation}
N=N_r
\left(\frac{\Delta_r(t_r)}{\Delta(t)}\right)^{1/n}
\left(\frac{t}{t_r}\right)^{(n+1)/n}.
\label{scaling}
\end{equation}
We note that one numerical reference experiment per $n$th-order algorithm is sufficient
to determine the parameters $N_r$, $\Delta_r(t_r)$, and $t_r$.
While these parameters may be different for different systems, the scaling of $N$ with
$t^{3/2}$ and with $t^{5/4}$, respectively, for second- and fourth-order algorithms, will not be affected.
Most importantly, since the number of matrix-vector operations required by
the one-step algorithm scales linearly with $t$, it is clear that for long enough
times $t$, the one-step algorithm will be orders of magnitude more efficient
than the current FDTD methods.
In Fig.\ref{scalingplot} we show the required number of operations
as a function of time $t$ taking, as an example, simulation data of 3D systems (discussed below)
to fix the parameters $N_r$, $\Delta_r(t_r)$, and $t_r$.
We conclude that for longer times
none of the FDTD algorithms can compete with the one-step algorithm in terms of efficiency.
For $t=20$, the one-step algorithm is a factor of ten faster than the Yee algorithm.
Thereby we have disregarded the fact that the Yee
algorithm yields results within an error of 0.1\% while the one-step algorithm
gives the numerically exact solution.

\begin{figure}[t]
\begin{center}
\includegraphics[width=16.cm]{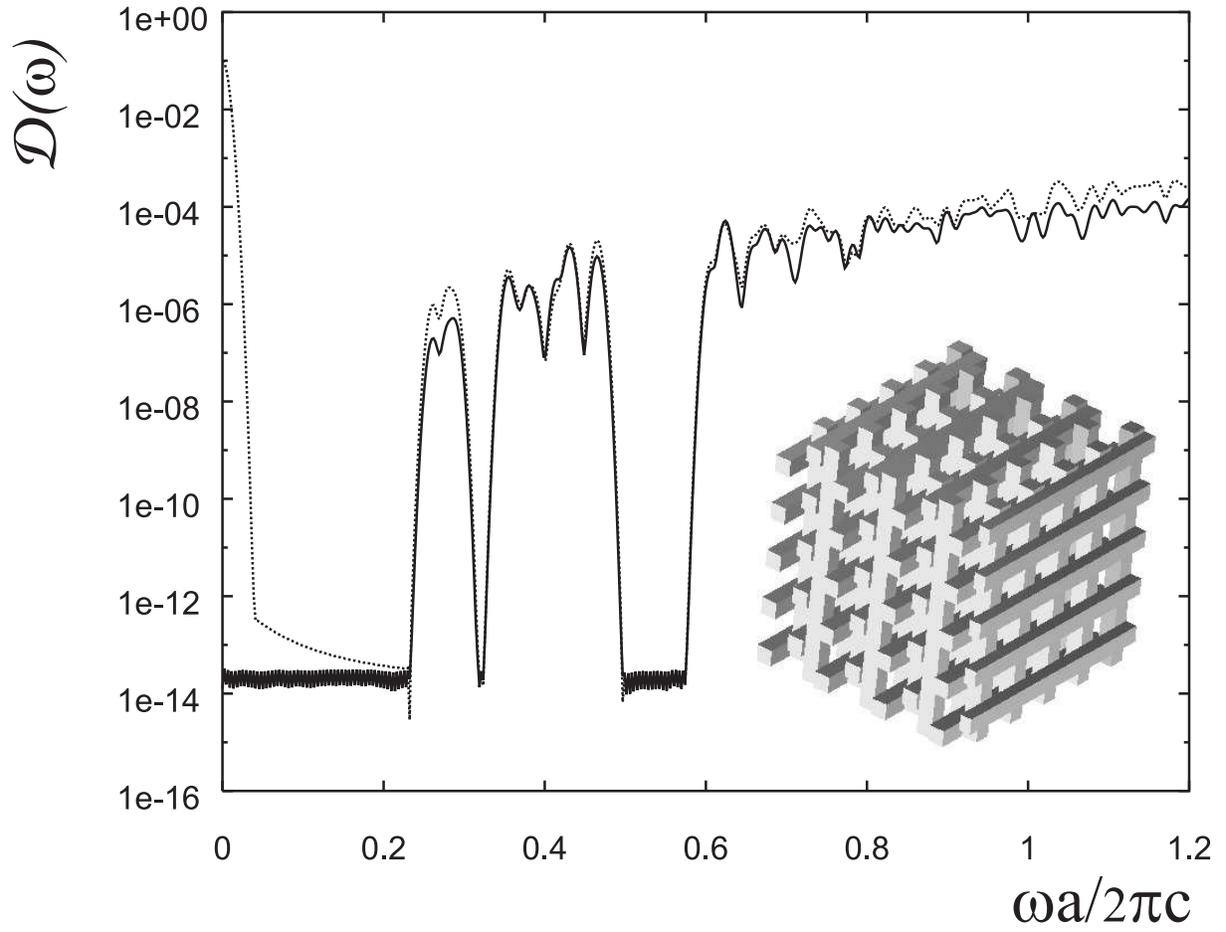}
%\unitlength 1cm
%\begin{picture}(2.0,2.0)(-2.5,-4.0)
%\includegraphics[width=2.cm]{woodpile2.eps}
%\end{picture}
%\vskip -1.5cm
\caption{Frequency spectrum of a three-dimensional photonic woodpile (inset)~\cite{Lin98}
as obtained by $T4S2$ (dashed line) and the one-step algorithm (solid line).
The width, height and period of the rods are 0.55, 0.7, and 2, respectively.
The dielectric constant of the rods is 12.96 and the simulation box measures
$6\times6\times5.6$~\cite{units}, subject to periodic boundary conditions.}
\label{woodpile}
\end{center}
\end{figure}

As the second example we use the one-step algorithm to
compute the frequency spectrum of a three-dimensional photonic woodpile~\cite{Lin98}.
This structure, shown in the inset of Fig.~\ref{woodpile},
possesses a large infrared bandgap and is under current experimental and theoretical
investigation~\cite{Lin98,Fleming02}.
To determine all eigenvalues of the corresponding matrix $H$
we follow the procedure described in Refs.~\cite{Kole01,Alben75,Hams00}.
We use random numbers to initialize the elements of the vector ${\mathbf \Psi}(0)$.
Then we calculate the inner product $F(t)=\langle{\mathbf \Psi}(0)|{\mathbf \Psi}(t)\rangle$
as a function of $t$ and average $f(t)=F(t)/F(0)$ over several realizations of the
initial vector ${\mathbf \Psi}(0)$.
The full eigenmode distribution, ${\cal D}(\omega)$, is obtained by Fourier transformation of $f(t)$.
In Fig.~\ref{woodpile} we show ${\cal D}(\omega)$, as obtained by $T4S2$
and the one-step algorithm,
with a time step $\tau=0.075$ (set by the largest eigenvalue of $H$),
a mesh size $\delta=0.1$, and 8192 time steps.
For this choice of parameters, the Yee algorithm would be unstable~\cite{Taflove,Kunz}
and would yield meaningless results.
The $T4S2$ calculation shows a peak at $\omega=0$. This reflects the fact that, in a strict sense,
the $T4S2$ algorithm does not conserve $\nabla\cdot(\sqrt{\mu}\bX(t))$ and
$\nabla\cdot(\sqrt{\ve}\bY(t))$~\cite{Kole01,Kole02}.
However, the peak at $\omega=0$ vanishes as $\tau^4$.
Repeating the $T4S2$ calculation with $\tau=0.01$ yields a ${\cal D}(\omega)$ (not shown)
that is on top of the result of the one-step algorithm (see Fig.~\ref{woodpile}) and is
in good agreement with band-structure calculations~\cite{Lin98}.
For $\tau=0.01$ the one-step algorithm is 3.5 times more efficient than $T4S2$.
Note that in this example, the one-step algorithm is used for a purpose
for which it is least efficient (time-stepping with relatively small time steps).
Nevertheless the gain in efficiency is still substantial.
In simulations of the scattering of the EM fields from the same woodpile (results not shown),
the one-step algorithm is one to two orders of magnitude more efficient than current FDTD algorithms,
in full agreement with the error scaling analysis given above.

\section{Conclusion}\label{sec4}
We have described a one-step algorithm, based on the Chebyshev polynomial expansions,
to solve the time-dependent Maxwell equations %for $d$-dimensional systems ($d=1,2,3$)
with spatially varying permittivity and permeability and current sources.
In practice this algorithm is as easy to implement as FDTD algorithms.
Our error scaling analysis shows and our numerical experiments confirm that
for long times the one-step algorithm can be orders of magnitude more efficient than current FDTD algorithms.
This opens possibilities to solve problems in computational electrodynamics that are currently intractable.

\begin{acknowledgments}
H.D.R. and K.M. are grateful to T. Iitaka for drawing our attention to
the potential of the Chebyshev method and for illuminating discussions.
%This work is partially supported by the Dutch `Stichting Nationale Computer Faciliteiten'
%(NCF), and the EC IST project CANVAD.
\end{acknowledgments}

\newpage

\begin{table}[t]
\begin{center}
\caption{The error $\Delta(t)$ after simulation time $t=100$ as a function of the time
step $\tau$ for two FDTD algorithms.
The number of matrix-vector operations required to compute the solution,
is $K^\prime=2080$, $t/\tau$, and $6t/\tau$
for the one-step, Yee, and $T4S2$ algorithm, respectively.}
\label{tab1}
\begin{ruledtabular}\begin{tabular}{ccc}
$\tau$ & Yee &  $T4S2$ \\
\hline
$0.10000\times10^{+0}$ & $0.75\times10^{-1}$ & $0.51\times10^{-1}$ \\
$0.50000\times10^{-1}$ & $0.25\times10^{-1}$ & $0.33\times10^{-2}$ \\
$0.25000\times10^{-1}$ & $0.12\times10^{-1}$ & $0.21\times10^{-3}$ \\
$0.12500\times10^{-1}$ & $0.66\times10^{-2}$ & $0.13\times10^{-4}$ \\
$0.62500\times10^{-2}$ & $0.24\times10^{-2}$ & $0.91\times10^{-6}$ \\
$0.31250\times10^{-2}$ & $0.63\times10^{-3}$ & $0.30\times10^{-6}$ \\
$0.15625\times10^{-2}$ & $0.16\times10^{-3}$ & $0.15\times10^{-7}$ \\
$0.78125\times10^{-3}$ & $0.39\times10^{-4}$ & $0.60\times10^{-8}$ \\
\end{tabular}
\end{ruledtabular}
\end{center}
\end{table}

\newpage

\newpage

\newpage

\begin{thebibliography}{99}
\bibitem{units}
We measure distances in units of {$\lambda$}.
Time and frequency are expressed in units of {$\lambda/c$} and {$c/\lambda$}, respectively.
\end{thebibliography}
\end{document}